\documentstyle[lscape]{mn}

\begin{document}

\title[Red Giant Branch Photometry II]{The Analysis of Isochrone Fitting
Methods for Red Giant Branch Photometry, and Tip Red Giant Branch Distance Determination}

\author[C. M. Frayn and G. F. Gilmore]
       {C. M. Frayn$^{1}$, G. F. Gilmore$^{1}$\\
       $^{1}$ Institute of Astronomy, University of Cambridge, Cambridge, CB3 0HA, UK\\
}
\date{Submitted to MNRAS, September 2002}

\pagerange{\pageref{firstpage}--\pageref{lastpage}}
\pubyear{200?}

\maketitle

\label{firstpage}

\begin{abstract}
We consider the procedure of isochrone fitting and its application to the
study of red giant branch (RGB) photometry in old stellar populations. This
is extended to consider the problems introduced by the inclusion of
lower-magnitude regions of the Colour Magnitude Diagram (CMD).  We refer
especially to our previous paper Frayn \& Gilmore (2002), where the details
of our isochrone interpolation and fitting code are explained.

We address the systematic errors inherent in the process of isochrone
fitting, and investigate the extent to which simple stellar populations can
be recovered from noisy photometric data.  We investigate the effects caused
by inaccurate distance estimates, isochrone model variation and photometric
errors.  We present results from two studies of approximately coeval stellar
populations, those of the Milky Way Globular Cluster System (MWGCS), and the
Ursa Minor dwarf spheroidal galaxy.

In addition, we introduce a new method for estimating distances using
photometry of the tip of the RGB which is significantly more robust than the
standard edge-detection filter.

\end{abstract}


\section{Introduction}

In our first paper, Frayn \& Gilmore 2002 (henceforth FG1), we introduced
isochrone-fitting methods to analyse photometric data on red-giant branch
populations in the haloes of nearby galaxies. Best-fitting metallicity
distributions were obtained by generating a probability matrix mapping each
star in a theoretical or experimental Colour-Magnitude Diagram onto an
occupation probability for each isochrone in the input model set. Isochrone
weighting coefficients were then optimised to produce a profile which
maximised the likelihood of producing the observed CMD under a set of
constraints. Most importantly, we considered only a high-age coeval
distribution, thus largely overcoming the problems of the age-metallicity
degeneracy. We refer the reader to FG1 for more details.

In FG1, we treated many sources of error in an attempt to gain the most
accurate metallicity distribution possible. However, it is clear that the
technique of isochrone fitting is extremely problematic, and many effects
need careful consideration if incorrect or overly optimistic conclusions are
to be avoided.

In this paper, we re-examine the most important sources of systematic error,
and investigate the effects that errors have on the conclusions we can draw
from RGB studies.  We also consider the effects of including deeper
photometry down to the Horizontal Branch (HB) or Main-Sequence Turn-Off
(MSTO) regions, and compare the results gained from analysis of the entire
CMD to those gained from considering the RGB only.

In order of importance, we investigate the primary causes of error that we
have discovered, concentrating on the methods outlined in FG1.  We identify
the errors inherent in those, and indeed any similar methods, in a
quantitative manner. As an appendix, we consider the uncertainties inherent
in the isochrones themselves.

We also present a new method for obtaining distance estimates from RGB
photometry that yields more accurate results than a traditional edge
detection implementation for the majority of cases, and is robust even to
severe photometric errors.

\section{Limitations due to the analysis method}

In this section, we deal with the problems caused by the fundamental
assumptions that we have to put into our population analysis, and
investigate the degree to which we can resolve these problems using
different methods.

  \subsection{Age-metallicity degeneracy}

This is the main problem that must be faced in the analysis of RGB stellar
populations.  As is well known, and reviewed in FG1, the effects of
increasing metal abundance on stellar isochrones are remarkably similar to
those of increasing age.  In fact, with the absence of any other diagnostic
methods, it becomes impossible to separate the two effects in some parts of
the CMD. This effect is particularly bad in the RGB which has no sharp
features to help distinguish between the contrasting effects.

We avoid this problem by assuming a high age, and then reducing the data
as if it were coeval.  This method is justified when dealing with primarily
old populations where the effects of age on the RGB become significantly
lessened (see FG1).  However, metallicity still has noticeable effects, and
therefore we can get a rough metal abundance distribution without having to
worry too much about the age spread.

\begin{figure}
\vspace*{7.2cm}
\includegraphics{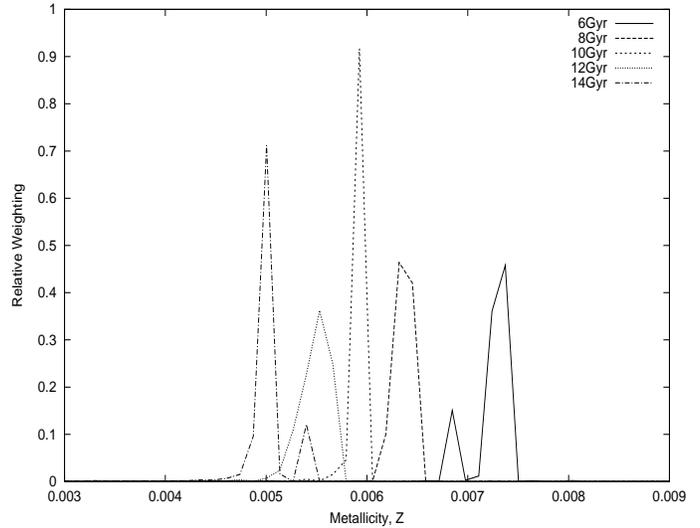}
\caption{A single delta-function RGB population at a metallicity of $Z =
0.006$ and age 10Gyr. This was then reduced assuming incorrect ages between
6 Gyr and 14 Gyr, i.e. $\pm 40\%$.  Although the error is noticeable, it is
not drastic, even at a large age error.  Note the shortened x-axis scale.}
\label{agezdeg}
\end{figure}

Here we test this hypothesis, investigating the degree to which the coeval
assumption in old RGB populations holds true.  We analyse synthetic CMDs
generated by our own code, and using the coeval fitting routines outlined in
FG1.  First of all we consider the pristine CMD, that is without any
photometric or Poisson errors added into the dataset.  We then reduce it
with only a tiny photometric error, essentially recovering the degree to
which the theoretical isochrones overlap.  This tells us something
fundmental about the actual limits to which we can trust isochrone fitting,
regardless of the data quality. Fig. \ref{agezdeg} shows the results of this
test.

The expectation is that we can only ever hope to achieve this quality of
data reduction at best on a true data set, simply because of the photometric
errors involved.  However, when attempting to recover a very narrow
distribution we may appear to manage slightly better.  This is simply
because, once accurate errors are considered, our maximum likelihood method
will attempt to narrow the metallicity distributions it finds if there is
insufficient data to suggest a substantial spread. In the absence of any
prior information to the contrary it is safe to assume that the stellar
populations are relatively simple.  In general, stars don't vary smoothly in
metallicity as a function of mass, which would be the effect caused by an
incorrect age assumption.

If one repeats the above experiment, but this time includes the full Poisson
errors in both the creation of the CMD and its analysis, and one also
incorporates fixed photometric errors, in this case we used 3\%, then the
reduction proceeds rather more smoothly.  The code has insufficient evidence
to suggest that the populations are anything other than delta-function in
metallicity, and assumes that the remainder of the spread is due to
photometric and Poisson errors.  Hence the resulting metallicity
distribution is significantly better.

\begin{figure}
\vspace*{7.2cm}
\includegraphics{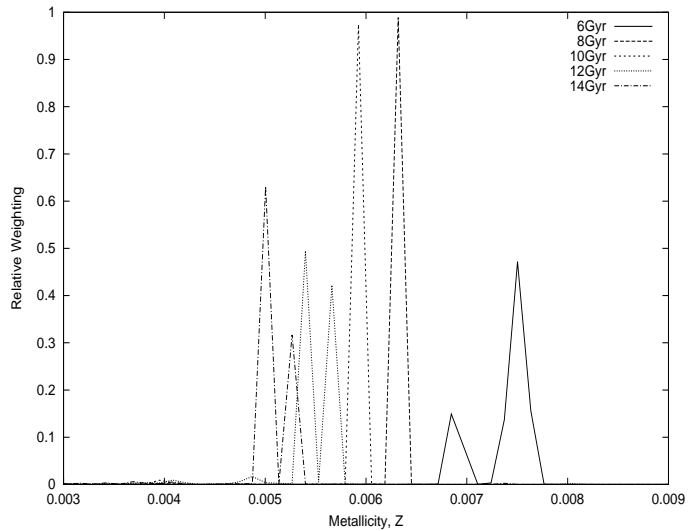}
\caption{A single delta-function population at a metallicity of $Z = 0.006$
and age 10Gyr.  The photometry was complete to beneath the MSTO. This was
then reduced assuming incorrect ages between 6 Gyr and 14 Gyr, i.e. $\pm
40\%$.  We introduced accurate photometric and Poisson errors into the
artificial datasets, and used the same values for our data analysis.  Using
accurate photometric errors helps in retrieving a more accurate guess to an
input metallicity distribution.}
\label{agezdeg2}
\end{figure}

These results are shown in Fig. \ref{agezdeg2}.  In this case we consider
photometry complete to the MSTO to demonstrate that even problems fitting
the HB are lessened once proper errors are considered.  This is an important
result -- HB contamination causes problems even when dealing primarily with
RGB photometry, if the completeness limit is deeper than the top few
magnitudes of the RGB.

Because of this inherent problem with our methods, we will encounter
problems with populations which are not truly of one unique metallicity, but
instead have some inherent spread in metallicity which should be recovered.
Often this spread is reduced due to the likelihood optimisation methods
employed. However, the perpendicular distance method (see FG1) seems to
recover input metallicity spread far more efficiently than a Gaussian
method.  In addition, underestimating the errors helps recover the spread,
albeit at a slight penalty in removing experimental noise.

  \subsubsection{Introducing age errors}

\begin{figure}
\vspace*{7.2cm}
\includegraphics{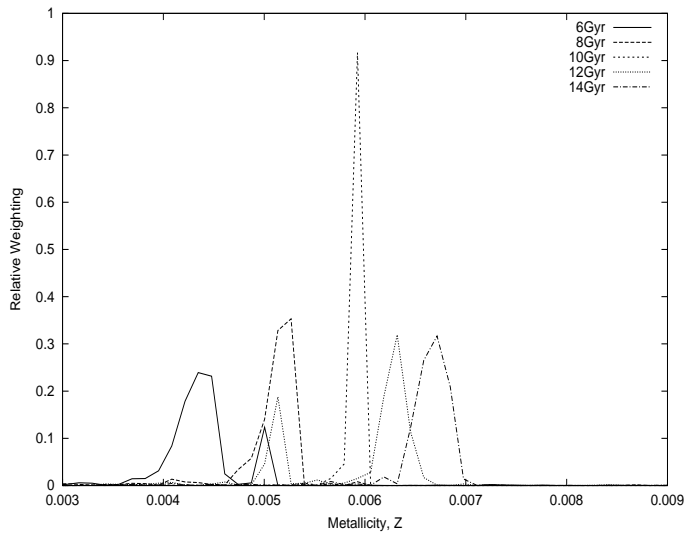}
\caption{A set of delta-function RGB populations at a metallicity of
$Z = 0.006$, but ages varying between 6 Gyr and 14 Gyr.  All were reduced under
the assumed age of 10 Gyr.  Errors are less severe for older populations.}
\label{varyage}
\end{figure}

Next we tested the reduction of populations generated at differing ages,
with an assumed age of 10 Gyr.  These results are shown in Fig.
\ref{varyage}.  Once again, the general form of the population is recovered,
though this recovery is obviously worse in the case of the younger
populations.

In all the cases investigated, the error becomes much greater when one
ventures to lower ages rather than high ages.  The difference between
populations of 10 Gyr and 14 Gyr is not too great, but the difference in
shape between 10 Gyr and 6 Gyr, though the same $\Delta age$, is far
greater.  At 6 Gyr, the isochrones include the effect of intermediate-age
Asymptotic Giant Branch (AGB) stars, and the change in isochrone slope
becomes greater per unit age interval.  Isochrone evolution is a roughly
logarithmic process, and most isochrone sets are logarithmically spaced in
age to account for this.

  \subsection{Distance degeneracy}

One further degeneracy problem must be considered in order to understand
fully the errors inherent in such a technique.  That is the principle of
metallicity-distance degeneracies.

So far we have considered populations located at an accurately-known
location.  We have considered that the distance modulus of the target
population is known precisely.  In fact, this is rarely the case.  For
distant targets, the distance might be unknown by a large percentage.  At
the Virgo cluster, the best guess distances to target galaxies are known
only to around the $\pm 0.2$ magnitude level (e.g. Ferrarese et al. (1996))
from Cepheid variable methods.  It is important to measure the extent of the
problems caused by assuming an incorrect distance modulus for a target
population.

In order to test this, we have generated an artificial CMD as described in
our previous paper.  We have generated a single delta-function
population at the distance of 16Mpc, consisting of 10,000 stars.  We then
reduced this population using our numerical techniques assuming the true
distance modulus of $+31 mag$, and then over- and under-estimating this
value by $0.1 mag$ and $0.2 mag$.  We considered only the RGB for analysis,
as before, using the 48 RGB stars thus obtained.

\begin{figure}
\vspace*{7.2cm}
\includegraphics{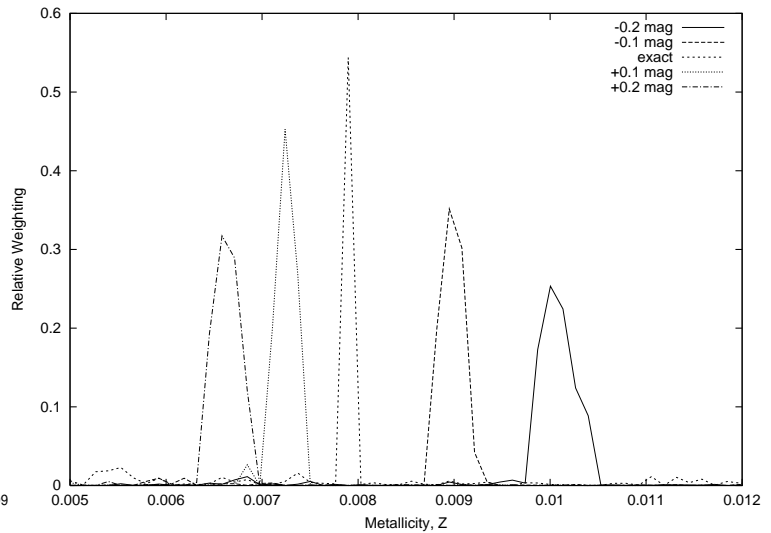}
\caption{A single delta-function population at a metallicity of $Z = 0.008$.
This was then reduced assuming slight errors in the distance modulus varying
between $-0.2$ and $+0.2$.}
\label{distanceerrors}
\end{figure}

Fig. \ref{distanceerrors} shows the results of this test.  An error of just
$\pm 0.2$ magnitudes in the estimated distance modulus can actually cause
appreciable problems with the recovered metallicity spread.  However, we can
account for this problem in our analysis of the recovered metal abundance
distributions.  Several methods can be used to estimate distances to nearby
galaxies, including Cepheid variable studies, and Tip-Red Giant Branch
(TRGB) photometry.

A distance overestimate of $+0.2 mag$ equates to an underestimate of the
true metallicity by approximately 20\%. Underestimating the distance by the
same amount equates to an overestimate of the true metallicity by 25\%.
Both error directions also cause a spread in the recovered metallicity
distribution.  Even a 25\% error in the recovered metal abundance is not an
enormous problem when considering the difference between an early, low
metallicity population and a late, high metallicity population whose metal
abundances might well vary by a factor of three or more.

 \subsection{Reddening effects}

Interstellar reddening affects the accuracy to which we can understand any
stellar population.  In the optical regime, dust causes a
wavelength-dependent absorption feature which decreases apparent
luminosities more strongly in the blue end of the spectrum than in the red.
Hence, the $V$ and $I$ luminosities are reduced, but to differing amounts,
therefore also altering the $V-I$ colours correspondingly.

\begin{figure}
\vspace*{7.2cm}
\includegraphics{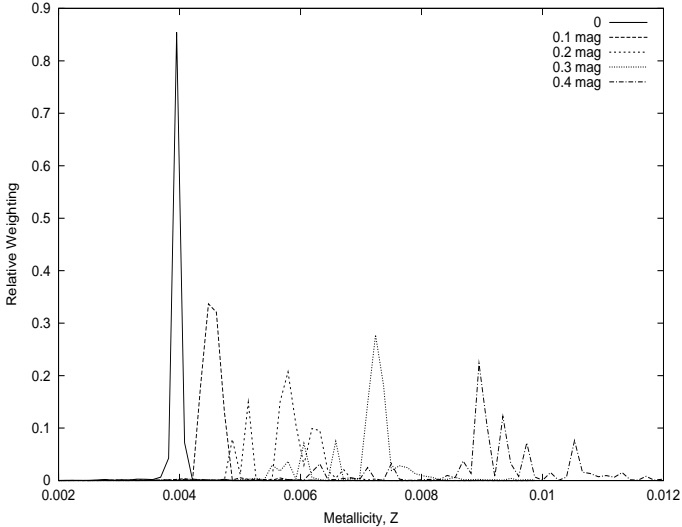}
\caption{A single delta-function population at a metallicity of $Z = 0.004$ and
age 10 Gyr which has been reddened by varying amounts, and then reduced
assuming no reddening whatsoever.  Reddening causes an overestimate of the
true central metallicity value and a spread in the distribution.  The five
plots represent no reddening, and then for $V$-band extinction of between
0.1 and 0.4 mags.  The effect on $V-I$ colours was taken from Mathis
(1990).}
\label{reddening}
\end{figure}

To test the actual effect of reddening on a simple stellar population, we
have run a series of simulations.  We applied varying levels of reddening to
a simple stellar population, and ran this through our code assuming no
reddening whatsoever.  The results are shown in Fig. \ref{reddening}.  We
assume the extinction model proposed by Mathis (1990).

Reddening causes two effects.  Firstly, it dims the stars in each optical
filter, simulating effects similar to a distance increase.  Secondly, it
shifts the colours of the stars redward, thereby mimicking a slight
metallicity increase.

\section{TRGB distance estimation}
\label{TRGBd}

We have shown that the assumed distance for a stellar population can indeed
cause several problems if it is significantly different to the true value.
It makes sense therefore to use our RGB photometry in order to estimate a
good fit to the distance modulus using the TRGB distance estimation
technique, outlined by Salaris \& Cassisi (1997) amongst others.

This method relies on the fact that the total bolometric magnitude of the
tip of the RGB appears to be approximately constant, regardless of age and
metallicity, for stable old populations.  The absolute $I$ magnitude of the
tip is also approximately constant, at a value of $I \simeq -4.05 mag$,
varying by no more than 0.1 magnitude for $-2.2 < [Fe/H] < -0.7$, according
to Da Costa \& Armandroff (1990).

We can use this method therefore to check the assumed distance modulus and
to verify that a sensible value has been used.  To do this, we require an
edge fitting method which will allow us to detect the apparent $I$ magnitude
of the TRGB from our data.

There is some discrepancy in the literature about the true position of the
`edge' in this situation.  Many people follow the method of Sakai et al.
(1996), henceforth S96, who use a discrete Sobel edge detection filter to
find the assumed tip.  We remain sceptical about this particular method as
it is easily swamped by experimental Poisson statistics, especially in the
low-number count regime. 

We tentatively adopt the Sobel edge-detection method in order to test the
method.  Taking the given distance modulus as a first guess, we generated a
binned histogram between $\pm 1 mag$ of this value using an adaptive binning
technique which ensures that the number counts in each bin remain above a
certain low threshold.

\begin{figure}
\vspace*{7.2cm}
\includegraphics{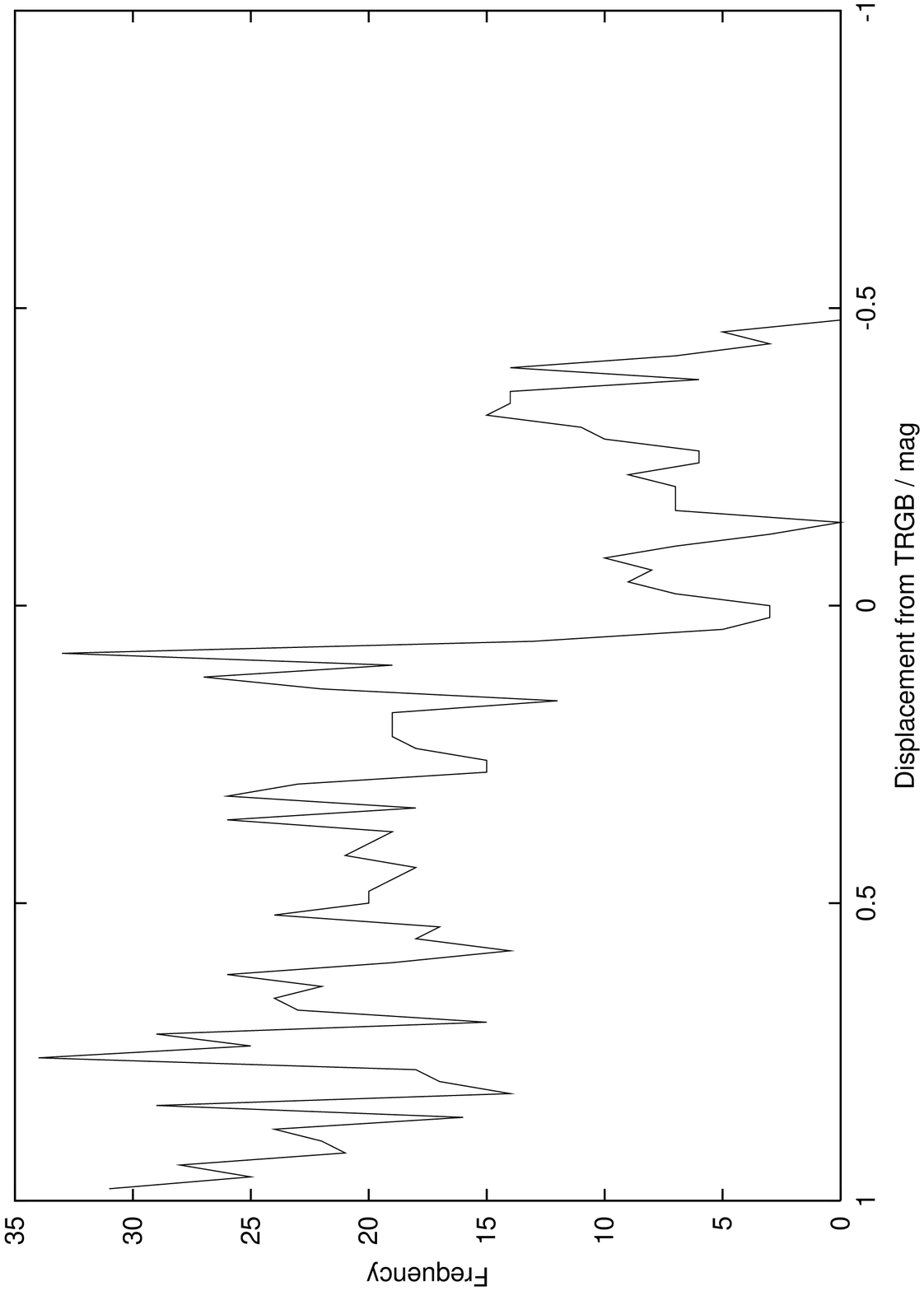}
\vspace*{7.2cm}
\includegraphics{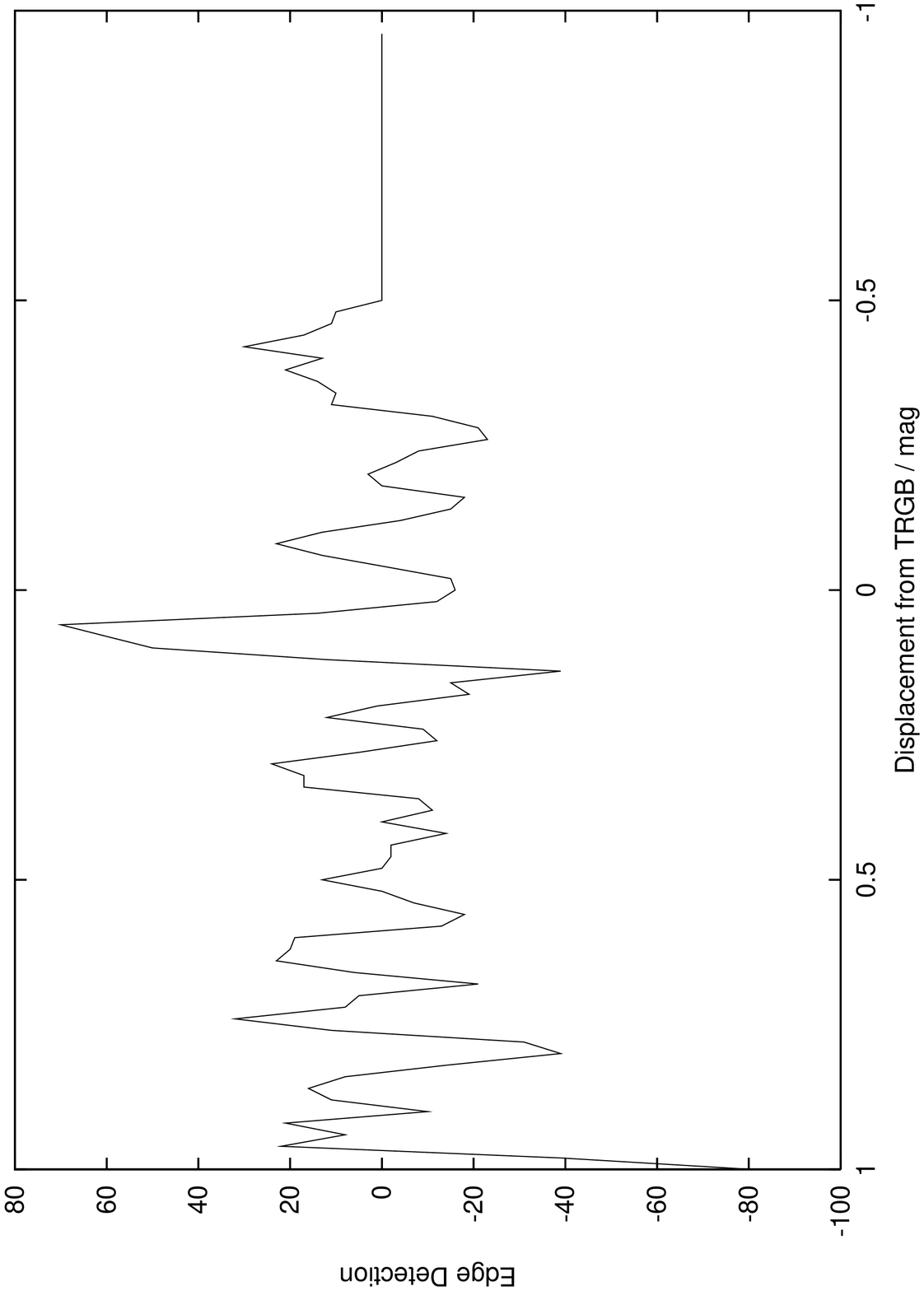}
\caption{An $I$-magnitude luminosity function for a coeval artificial
population at an age of 10Gyr with a central metallicity of one third solar
and an 8\% Gaussian spread.  No errors of any kind have been added to the
dataset, showing far more clearly the location of the TRGB edge.  Here,
displacement is measured in magnitudes from the theoretical value,
uncorrected for metallicity, of $I = -4.05$ (Da Costa \& Armandroff 1990).
In the second figure we present the results of a Sobel filter applied to
this TRGB luminosity function.  A clear edge is detected at a deviation of
0.06 magnitudes beneath the mean experimental value.}
\label{clean_edge}
\end{figure}

An example of the theory behind this method is illustrated in figure
\ref{clean_edge}.  Here we have generated a large artificial population
using the methods outlined in FG1, with a simple age and metallicity profile
and no photometric errors.  We present a binned $I$-magnitude luminosity
function for one magnitude either side of the expected RGB tip. An `edge' is
clearly seen.

Here we use the value of $I_{TRGB} = -4.05$ (Da Costa \& Armandroff 1990).
Values close to this seem to be preferred by most studies.  See e.g.
Salaris, Cassisi \& Weiss (2002).  Most authors seem to derive values
slightly brighter than $I_{TRGB} = -4.0$, though there is a spread of around
0.2 magnitudes in the literature.  We acknowledge that Girardi et al. (2000)
predict $I_{TRGB} \simeq -3.95$ from the theoretical isochrone models used
in this paper, which agrees slightly better with the position of the edge
determined in figure \ref{clean_edge}. Further work in this paper considers
only relative displacement so inaccuracies in this value are not important.

In more realistic stellar populations, there are two additional spreads in
the $I$-luminosity function.  One is the intrinsic spread due to the fact
that not all of the stars are of the same metallicity, age and mass, in
general. However, there is also an extra spread caused by measurement errors
which makes this edge-detection method extremely difficult to apply.  Even
without measurement errors, one would not expect such a distinct edge as
that seen in Fig. \ref{clean_edge} in most populations, simply because they
are not coeval.  We consider an alternative : to calibrate the TRGB
distances using an appropriate Gaussian cutoff, and then search for the
mean.  This is much less susceptible to experimental errors and gives an
obvious target to identify.

First, we bin the data as above.  Then we fit a Gaussian cutoff to the
luminosity function of the form;

\begin{equation}
\begin{array}{lll}
\Phi = & A e^{(\frac{-(m_I-b)^2}{2c^2})} + d & for \ m_I > b\\
\Phi = & A + d & otherwise\\
\end{array}
\label{TRGBdist}
\end{equation}

and solve for the best-fitting values for the four parameters.  $A$ gives a
scaling value, $b$ gives an $I$-magnitude shift, $c$ gives a scale length
for the TRGB tail, and $d$ gives a constant background offset.

Once these four parameters have been found, then we require the
$I$-magnitude of the maximal gradient in the luminosity function, or
alternatively, the point where the second derivative is zero. Analytically,
this is at the value $I = c + b$, which can then be calculated from the fit
parameters.  This gives us a good estimate for the TRGB magnitude, which can
be compared with the value taken from calibration on nearby globular
clusters with well-known distances.  Clearly, we could choose any such point
on the Gaussian curve and calibrate our relation with comparison to
theoretical datasets, but this seems the easiest to define.

In order to compare the two contrasting methods, a set of 16 large CMDs was
produced using our code.  Each of these was then analysed using both the
traditional Sobel edge detection method, and also by fitting a Gaussian
cutoff of the form shown above in equation \ref{TRGBdist}.  For each dataset
we used exactly the same initial conditions, that is a single population
with an age of $10 \pm 0.2 Gyr$ and a metallicity of $Z = 0.006 \pm 0.0005$.
A Salpeter IMF was used, and the Girardi isochrone set.  Realistic
photometric errors of 6\% were added.

\begin{figure}
\vspace*{7cm}
\includegraphics{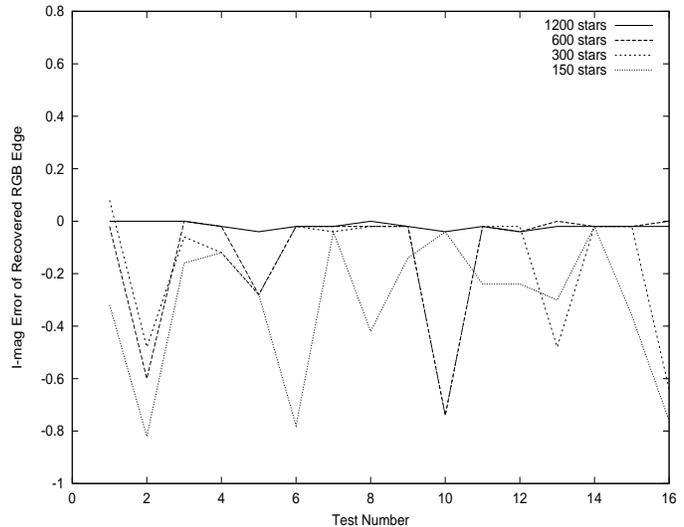}
\caption{Error ($I_{true} - I_{detected}$) on the RGB tip determination using
a Sobel filter method like that proposed by S96.  The population is a simple
stellar cluster at $10 \pm 0.2$ Gyr and $Z = 0.006 \pm 0.0005$.  IMF is
Salpeter, and photometric errors of 6\% were added. 16 test populations were
used. Positive values represent an overestimate of the TRGB magnitude.}
\label{TRGBsobel}
\end{figure}

\begin{figure}
\vspace*{7.2cm}
\includegraphics{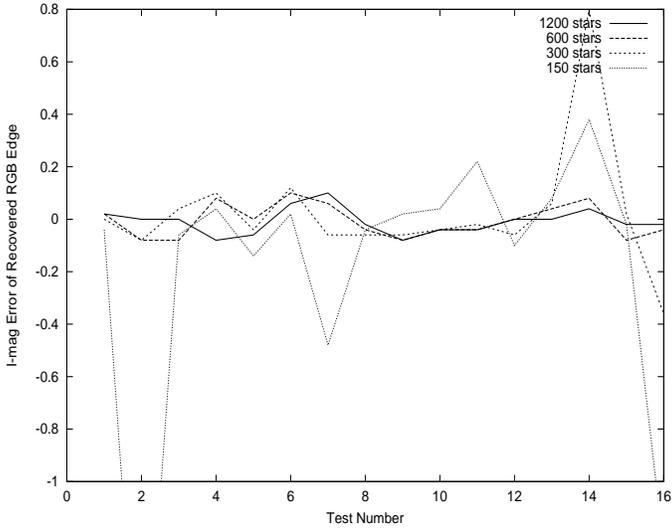}
\caption{Error ($I_{true} - I_{detected}$) on the RGB tip determination using
a Gaussian fitting method as described in this section, for the same
population described in figure \ref{TRGBsobel}.  Errors are much smaller in
general, except for the occasional large discrepancy caused by a bad fit - a
feature that is very easy to test for and avoid. Positive values represent
an overestimate of the TRGB magnitude.}
\label{TRGBgauss}
\end{figure}

In each set the initial TRGB detection was approximately 1200 stars within
one magnitude of the expected TRGB luminosity in $I$.  Next, half of the
stars were removed at random in three steps, thereby creating three further
datasets with 600, 300 and 150 stars within one magnitude of the TRGB in
each respectively.  The same statistical process was repeated on each
dataset.  Then we repeated the TRGB detection with each dataset after
reduction.  The results are plotted in figures \ref{TRGBsobel} and
\ref{TRGBgauss} for the edge-detection and Gaussian cutoff methods
respectively.

What is clear is that the scatter on the values for a Sobel edge fit is
smaller for the initial dataset, with approximately 1200 stars within one
magnitude of the TRGB. However, as soon as the number of detections reduces
significantly beneath 1,000, the Gaussian method produces more reliable
results.  The one caveat is that, by the nature of the method, the Gaussian
technique occasionally produces wildly inaccurate results.  This is always
easy to detect as the fit Chi-squared value will be very poor, and the
values for the parameters $A,b,c$ and $d$ will lie outside realistic ranges.
For the Sobel method, it is impossible to tell $a posteriori$ which results
are inaccurate.

Removing those results with clearly poor fits, the average statistical
deviation from the true TRGB edge value was calculated in each case of
1,200, 600, 300 and 150 stars within one magnitude of the TRGB. We measured
absolute displacement from the mean location of the edge obtained from the
largest populations, thus automatically calibrating the Gaussian method and
avoiding any problems with the precise magnitude of the TRGB edge used for
the Sobel method.  For the Sobel filter method, the values were $\sigma =
0.02, 0.25, 0.31, 0.40$ magnitudes respectively. This means that even with
150 stars within one magnitude of the TRGB, the expected error on the TRGB
detection is 0.4 magnitudes, and could well be much worse.

For the Gaussian method, these values were $\sigma = 0.05, 0.06, 0.06, 0.14$
magnitudes - an improvement of a factor of almost three on the worst case.

\begin{figure}
\vspace*{7.2cm}
\includegraphics{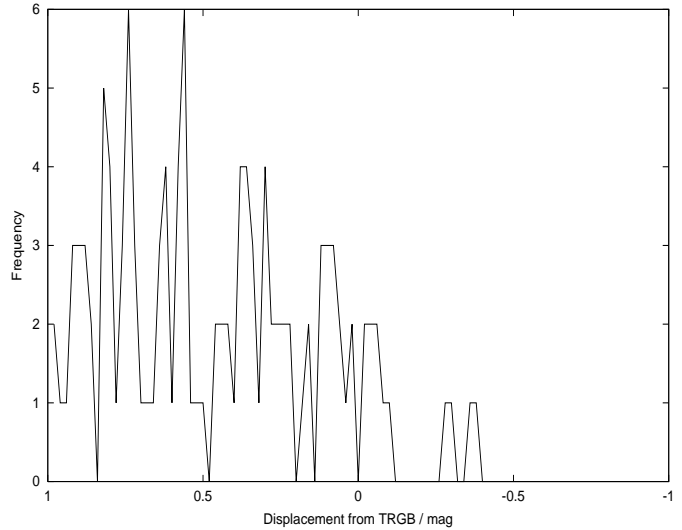}
\caption{The TRGB luminosity function for the Ursa Minor dSph, showing a
binned histogram for the 121 stars within one magnitude of the theoretical
TRGB.  Clearly low-number statistics would make any distance determination
using a discrete Sobel filter completely meaningless.  Fitting a Gaussian to
the results however, as described in this work, gives a distance estimate
different from the theoretical value (Mateo, 1998) by just 0.04 magnitudes.}
\label{UMi_TRGB}
\end{figure}

For most datasets we do not expect a substantial number of stars near the
TRGB magnitude.  Certainly we expect numbers in the low hundreds at best. An
example is shown in Fig. \ref{UMi_TRGB}, showing the TRGB luminosity
function for stars in the Ursa Minor dataset analysed in section
\ref{UMi_analysis}. Clearly, any edge detection on data with such low-number
statistics will be largely random.  However, by fitting the TRGB with our
Gaussian model, as described in this section, we obtain a distance estimate
which is only different from the theoretical value of $(m-M) = 19.11$
(Mateo, 1998) by $+0.04$ magnitudes.  We note that the theoretical value was
based on studies of variable stars, so this acts as an independent
justification for our techniques.

 \subsection{Varying photometric errors}

\begin{figure}
\vspace*{7.2cm}
\includegraphics{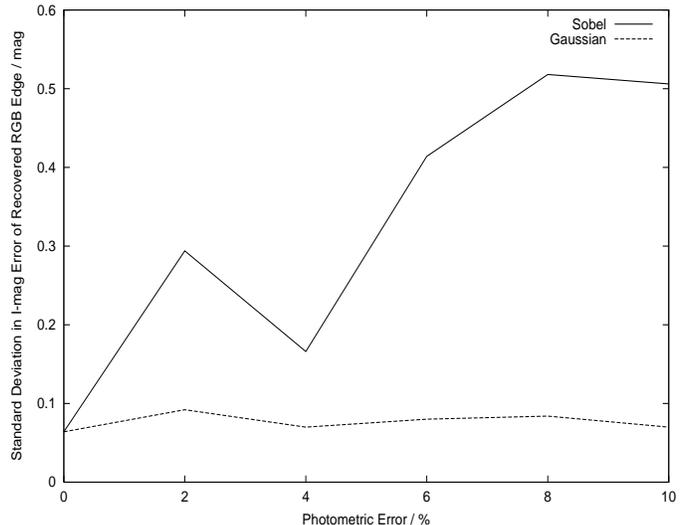}
\caption{Testing the new Gaussian method for estimating the TRGB magnitude
in simulated stellar populations.  Varying photometric errors have been
used.  Population details remain the same as that in figure
\ref{TRGBsobel}.  Clearly the Gaussian method remains superior at all levels
of photometric error.}
\label{VarPerr}
\end{figure}

In order to test this method further, we created a set of datasets identical
to those above, but with varying photometric errors between 0\% and 10\%.
These datasets were reduced using the Sobel filter method and the Gaussian
method.  The population was created so that approximately 500 stars were
found within one magnitude of the TRGB in the each dataset, which is more
than could be obtained for most observational datasets. The performance of
the Sobel method in this test is therefore slightly {\em better} than might
be expected in general.  Four populations were created for each photometric
error value, and the standard deviation in the errors from the true TRGB
location was calculated for each method.  The results are shown in Fig.
\ref{VarPerr}.

As the results clearly show, the Gaussian method remains preferable over the
Sobel filter method at all photometric error levels.  It is hardly affected
at all by the introduction of severe photometric errors, as one would expect.

 \subsection{Conclusions}

We therefore suggest a new method for estimating TRGB distances in average-
to poorly-populated CMDs by fitting the above proposed Gaussian cutoff
(formula \ref{TRGBdist}) to the TRGB luminosity function and solving for a
fixed point on the curve, such as the zero of the 2nd differential at $I = c
+ b$.  We propose that this method is robust to numerical errors, and allows
an accurate determination of the TRGB distance especially for populations
with a poorly-defined RGB tip.

For larger datasets, that is if there are at least 1000 stars within one
magnitude of the TRGB, the standard edge-detection method seems more robust,
though we instead suggest adoption of the technique outlined by Cioni et al.
(2000) using the second derivative of the luminosity function. They report
that this method works more reliably than the traditional Sobel filter
method, though obviously it will still suffer from the above described
numerical errors with smaller datasets.

We now present an application of the current work to real astronomical
datasets in order to demonstrate its effectiveness and accuracy.  We present
two different datasets in order to test the robustness of our code in
different situations.

\section{Milky Way globular clusters}

We obtained a set of 20 CMDs from the publically available archive of
Rosenberg et al. (2000) and tested our code on all of them.  One of the most
important effects we noticed was the inability for our code to fit very
metal-poor populations, due to isochrone limits.  This permits us only to
obtain an upper limit on the true metallicity.

\begin{figure}
\vspace*{7.5cm}
\includegraphics{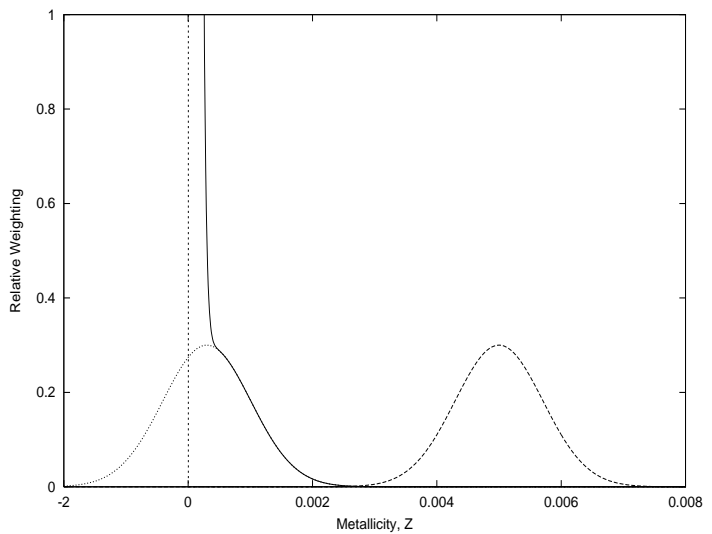}
\caption{The minimum metallicity we can accurately detect is approximately
$Z = 0.0006$, due to the limits of the weighting scheme we employ.  At higher
metallicity (rightmost peak) we detect the peak accurately.  When the
metallicity gets very low (leftmost peak) then every metallicity value lower
than our lowest-metallicity isochrone is added together into the
lowest-metallicity weighting value.  The dotted line shows the true shape,
and the solid line shows the recovered profile.  Because of this effect, it
becomes impossible to detect peaks accurately if the true mean falls within
the first three or four isochrone metallicity values.}
\label{upperlim}
\end{figure}

The reason why we obtain only a limit on the metallicity is simply because
we don't have any isochrones beneath $Z = 0.0004$.  We run into problems
with the weighting profiles as soon as we get near this limit, as the
following diagrams show (see Fig. \ref{upperlim}), due to the fact that all
the weighting for metallicities lower than $Z = 0.0004$ is added to the $Z =
0.0004$ isochrone, as this is the nearest to the correct value.  It is
impossible to obtain a true value for metallicities lower than about $Z =
0.0006$ because of this shortfall.  The exact value depends, of course, on
the metallicity interpolation granularity in the isochrone set.  As a limit,
after some testing, we determined that the lowest metallicity peak we could
accurately detect was approximately that of the 4th most metal-poor
isochrone.

Of course, in reality values of $Z<0$ are not possible, so there is
actually a fundamental limit to which we can obtain metallicity peaks by
this method, regardless of isochrone spread.  At these levels we are running
into clear problems with our initial assumption that most star formation
events produce stars with a roughly Gaussian metallicity distribution.

Of all those datasets we managed to fit, after reddening correction and the
inclusion of the best literature values for the age and distance modulus to
each cluster, all but 3 were within 0.3dex of the Harris (1996) value for
the central metallicity.  Some of the variation in our results from the
accepted values may well be due to the inaccuracies in converting
metallicity abundances into $[Fe/H]$.  The three badly-fitting clusters were
NGC 5927, NGC 104 (47Tuc) and NGC 6205.

We also calculated the errors between the retrieved metallicity estimates
given the two assumed ages 10 Gyr and 13 Gyr.  We found that the estimated
absolute value of the discrepancy was $[Fe/H] \simeq 0.085 dex \pm 0.020
dex$ for our sample.

\begin{figure}
\vspace*{7.5cm}
\includegraphics{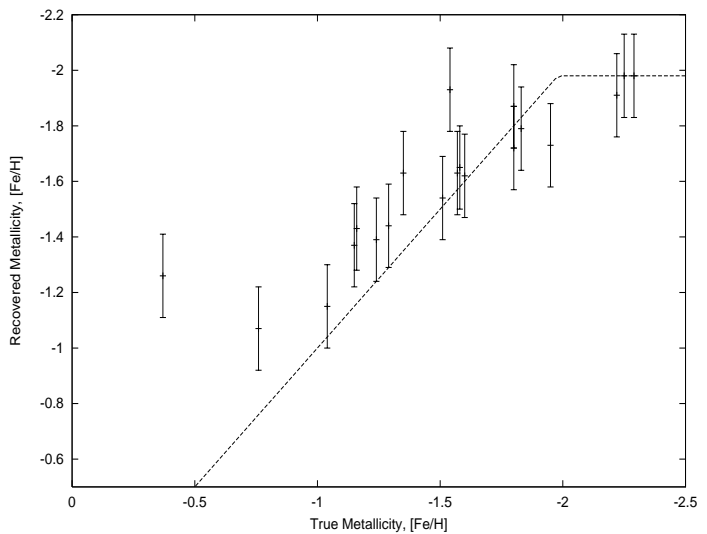}
\caption{Comparing our metallicity fits to Milky Way Globular Clusters with
the literature values of Harris (1996). Error bars are 0.15 dex, calculated
as the sum of the estimated uncertainties due to distance and age examined
above.}
\label{clusterfits}
\end{figure}

\section{Ursa Minor dwarf spheroidal galaxy}
\label{UMi_analysis}

The Ursa Minor dwarf spheroidal galaxy is a satellite galaxy to the Milky
Way, located at approximately 66 kpc from the sun (Mateo, 1998, henceforth
M98).  It is known to be a relatively simple stellar population, containing
mainly old stars, and possessing a narrow abundance distribution.  M98
suggests a single, large star formation between 10-14 Gyr ago, which
produced all the stars within the galaxy.  Ursa Minor has since remained
untouched by outside influence, except for the inevitable tidal disruption
caused by the Milky Way's gravitational potential.

As such, UMi is an interesting and instructive target with which to test our
code, knowing that we don't expect a purely coeval stellar population, but
that the contamination from young stars is low to negligible.  It provides a
suitable branch from the simplistic, coeval globular clusters to the more
complicated, spatially and chronologically extended dwarf members of the
local group.

  \subsection{The data}
  \label{UMidata}

We took our data from the photometry of Kleyna et al. (1998) who present a
Johnson-Cousins $V,I$ mosaic survey of the entire central region of the UMi
dwarf spheroidal to $V \simeq 22$, taken using the F. L. Whipple Observatory
(FLWO) 1.2 m telescope.  These data consist of 27 fields, imaged twice in
each of $V$ and $I$ for 300s per exposure.  Full details of the preliminary
data reduction can be found in Kleyna et al. (1998).

We obtained the full photometry dataset for all objects detected in the 27
fields, together with an estimate of the stellarity of each source as
calculated by the SEXtractor programme (Bertin \& Arnouts, 1996).  Non-stellar
objects were removed by imposing a selection criterion of $stellarity >
0.7$.  Further cuts were made in both absolute $I$ magnitude and $V-I$
colour in order to remove as large a fraction of contaminant foreground
stars as possible.  Stars with $I < 19.0$ were removed, together with all
stars outside the range $0.75 < V-I < 1.6$.  This left the RGB stars,
together with a significant amount of contaminants surrounding them. This
would therefore prove an excellent test of the ability of our {\it
FITCOEVAL} programme to ignore contaminants and pick out principal stellar
components from amongst noise.

\begin{figure}
\vspace*{7.5cm}
\includegraphics{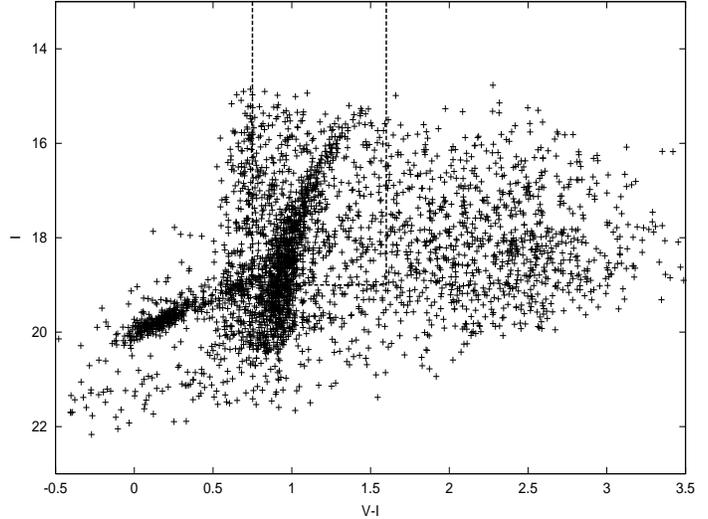}
\caption
{A CMD for the 4,370 stellar objects selected from the original UMi
photometry.  The solid lines show the cuts in both $I$ magnitude and $V-I$
colour imposed in order to select only the RGB population with as little
contamination as possible.  1021 stars passed the selection criteria.}
\label{UMi1}
\end{figure}

Fig. \ref{UMi1} shows the complete CMD for stellar objects in UMi,
together with the imposed cuts which we employed to select only the RGB
population.

\begin{figure}
\vspace*{7.5cm}
\includegraphics{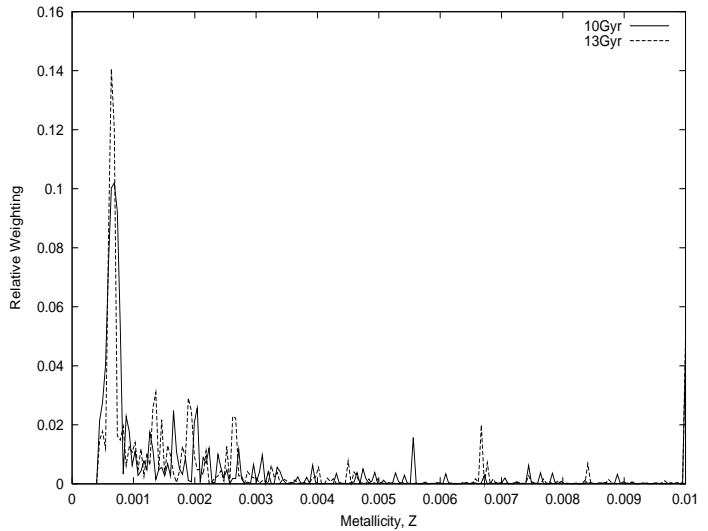}
\caption{Full reduction of the UMi data using an assumed distance of
$(m-M)_V = 19.11$, and isochrones of 10 Gyr and 13 Gyr.}
\label{UMi2}
\end{figure}

After obtaining the RGB population, we then added a reddening correction
using the literature value of $E(B-V) = 0.03 \pm 0.02 mag.$ Following this,
we ran our code on these data using the two spanning isochrones of 10 Gyr
and 13 Gyr.  Each of these were taken from a {\it finez} set, interpolated
to 200 metallicity subdivisions between $Z = 0.0004$ and $Z = 0.01$.  The
assumed distance was $(m-M)_V = 19.11$, from M98.  The optimum fits
calculated are shown in Fig. \ref{UMi2}.

The maximum values for each of these fits were $Z = 0.00069$ for the 10 Gyr
isochrone fit, and $Z = 0.00064$ for the 13 Gyr isochrone fit.  These
correspond to iron abundances of $[Fe/H] \simeq -1.79$ and $-1.82$
respectively.  Standard values for UMi place the iron abundance at roughly
$[Fe/H] \simeq -2.1$. (See, for example, M98).

As calculated earlier, we can estimate an error on this value of
approximately $0.15 dex$ from the uncertainties in distance and age.  The
uncertainty in the reddening is small ($0.02 mag$).  Using the lower value
of $E(B-V) = 0.01 mag$ alters the derived $[Fe/H]$ value by less than $0.01
dex$ in $[Fe/H]$.  Choosing a higher value of $E(B-V) = 0.05$ results in a
much poorer fit with a significantly lower final likelihood.  It provides a
pair of maxima in the resultant metallicity distribution around $Z = 0.00064$
and $Z = 0.00088$, corresponding to $[FeH] = -1.82$ and $-1.68$ respectively.

\section{Conclusion}

Our code appears reasonably robust to sensible photometric errors, and is
able to recover the central metallicity of a globular cluster stellar
population in most (75\% of) cases.  Resolution is restricted to a lowest
metallicity of $[Fe/H] \simeq -1.9$ due to numerical considerations.  At
lower metallicity, resolution in $[Fe/H]$ is reduced due to the logarithmic
nature of the iron abundance scale.

Age, reddening and distance errors should not be a great problem unless they
vary by a significant value, where `significant' here means $2-3 Gyr$ in
age, $0.1 mag$ in $V$-extinction or $0.2 mag$ in distance modulus.

C.M.F. would like to extend his thanks to Jan Kleyna for help in the
preparation of the code required for this project, and also to Lynette Dray,
Brian Chaboyer and Richard de Grijs for interesting discussions on these
topics.

\newpage

\section{Appendix : Difficulties with isochrones}

We briefly summarise some possible sources of systematic error in an adopted
set of isochrones.

In using the process of isochrone fitting, it is important first to
investigate the uncertainties caused by the isochrones themselves.  As
theoretical models, they have a considerable number of parameters whose
values can be changed, often causing significant differences between
different isochrone models.

A slight variation in the model definition at some rapid evolutionary stages
can give a drastically different isochrone. See, for example, the early
work done on convective overshooting by Bertelli et al. (1990).  The
isochrone set that we have mainly adopted for use in this study is that
by Girardi et al. (2000).

In general, the HB is very poorly understood in the theoretical models, with
little consensus on the true nature of the second parameter problem, and few
groups agreeing well on the predicted shape.  The horizontal branch occupies
a relatively compact area of the CMD, where small errors in photometry can
give very large errors in derived metallicity and hence errors in either the
models or the photometry in this regime. None of the isochrone sets we have
tested could accurately reproduce very blue HBs, such as that exhibited by
the galactic globular cluster NGC 288, amongst others.

Furthermore, if we continue the photometry even deeper then we approach the
main-sequence turn-off, where problems are caused by blue straggler stars
which are obviously not fitted at all well by any existing isochrone models,
being creations of stellar dynamics rather than isolated stellar evolution.
Binary stars can also cause problems, with theoretical isochrones only
representing single stellar systems.

  \subsection{Problems modelling the RGB}

The red-giant branch poses many unsolved problems for theoretical modellers,
and is consequently fairly inaccurately represented in most theoretical
isochrones. One such important problem is that of mass-loss.  RGB stars are
highly expanded with only tenuously retained outer layers.  Stars at this
stage in their evolution tend to lose a substantial fraction of their mass
through stellar winds.

Fortunately for the purposes of this study, mass loss on the RGB is only a
very minor effect for low-mass stars.  Provided it is treated sensibly, it
can almost be neglected.  Certainly for stars less than one solar mass, this
will not affect the isochrone models to any significant degree.

In addition, stellar atmospheres are not understood particularly well on the
RGB, leading to slightly larger colour-errors in these sections of the
models.  This affects all mass-ranges.  If one uses Kurucz model atmospheres
then the deviation from accuracy towards the faint end of the RGB can be
significant.  There is not a great deal that can be done to rectify this
problem, and the safest assumption seems to be to reduce the weighting for
stars near the tip of the RGB by introducing slightly larger underlying
intrinsic errors in this regime.

Derived values of the TRGB level vary by around 0.2 magnitudes between
individual theoretical studies.  This uncertainty is primarily caused by
uncertainties in the RGB modelling schemes used, and illustrates why
isochrone fitting must therefore remain an unreliable method for obtaining
absolute, quantitative values for the internal parameters of stellar
populations to any useful degree of confidence.

Finally, as mentioned in section \ref{mlt} below, mixing length theory and
the choice of surface boundary conditions can cause substantial variation of
the RGB model predictions. The reader is referred to sections 5.6 and 5.7 of
the review by Salaris et al. (2002) which covers these topics in
considerable detail.

  \subsection{Tuneable parameters}

Inherent in the theoretical models used to generate model isochrones, there
are a number of poorly-understood parameters whose values may be tuned in
order to generate the required results.  Often these parameters are altered
so that the isochrone fits a certain realistic CMD, but this is not always
the case.

  \subsubsection{Equation of state and opacities}

Adopted stellar opacity laws are continually being updated.  Corrections
from modern opacity calculations are rather smaller than could be expected a
generation ago, but they are still considerable under some circumstances.
Several studies have been performed on the effects of varying the opacity
and equation of state used in calculations, see for example, Yildiz \&
Kiziloglu (1997).

Rogers \& Iglesias (1996) have extended the standard OPAL opacity laws down
to lower stellar masses beneath $0.8 M_{\odot}$, therefore allowing more
accurate modelling of RGB stars in truly old, metal poor stellar
populations.  The life-span of an $0.8 M_{\odot}$ star is approximately one
Hubble time however, so these low-mass stars will not yet have reached the
RGB stage in most cases.

  \subsubsection{Mixing length theory \& convective overshooting}
  \label{mlt}
  
Mixing length theory comes from models of convective stars, and consequently
applies strongly to younger, more massive stars with convective cores.
However, it is also used in predicting the behaviour of the convective
envelopes of low-mass stars, to determine the temperature gradient in
superadiabatic regions. Hence the mixing length parameter determines the
effective temperature of stars with convective envelopes, such as the RGB
stars in which we are primarily interested.

For more massive stars, generally those above $1.6 M_{\odot}$, the free
parameter of the overshooting distance also becomes important. This was
treated in considerable detail by Maeder \& Meynet (1991).  They claimed
that including convective overshooting in their models could revise
estimated ages of Red Turnoff stars upwards by a factor of between 1.5 and
2.7.  Estimating the ages from the Blue Turnoff gave errors in age
determinations of factors between 1.6 and 2.2.

All isochrone models now include these parameters, but the actual extent to
which the convective overshooting model is applied, and the value for the
mixing length are still very much open to debate, and can actually cause a
substantial difference in the model colours obtained.

  \subsubsection{$\alpha$-element ratios}

Salaris, Chieffi \& Straniero (1993) examined in detail the degree to which
alpha-enhanced isochrone models alter the fits obtained to the galactic
globular cluster system and concluded that evolutionary properties of
stellar models depend significantly on the adopted abundances of all the
$\alpha-$elements.  They also discovered that adopted opacity laws can have
a substantial effect on the shape of an isochrone, especially at low
metallicity and low temperature.

\begin{figure}
\vspace*{7.2cm}
\includegraphics{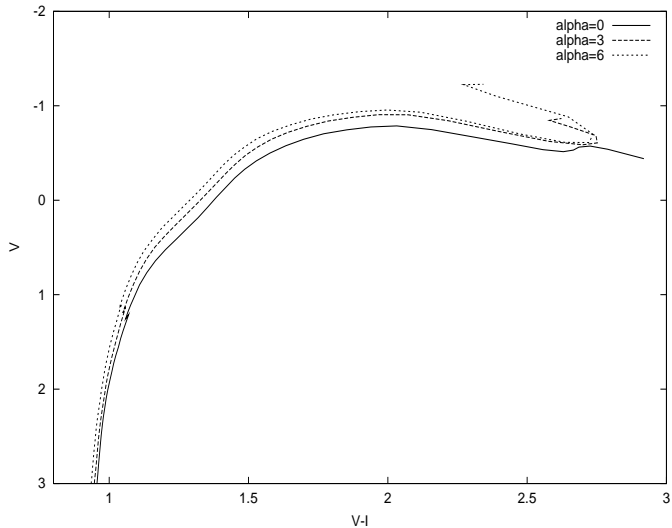}
\caption
{Here we have plotted isochrones for 10 Gyr with metallicity of one half solar
abundance, varying the alpha element ratios to $[\alpha/Fe] = 0.0, 0.3,
0.6$.  The difference is not enormous, but could cause a shift in the
assumed metallicity for RGB populations.}
\label{alpha}
\end{figure}

Recently, the Yale-Yonsei group have released a set of theoretical isochrone
tracks with varying alpha-element ratios (Yi et al., 2001).  Figure
\ref{alpha} shows the differences between alpha-element ratios of
$[\alpha/Fe] = 0, 0.3, 0.6$. Interpolation code kindly donated by Sukyoung
Yi.

\begin{figure}
\vspace*{7.2cm}
\includegraphics{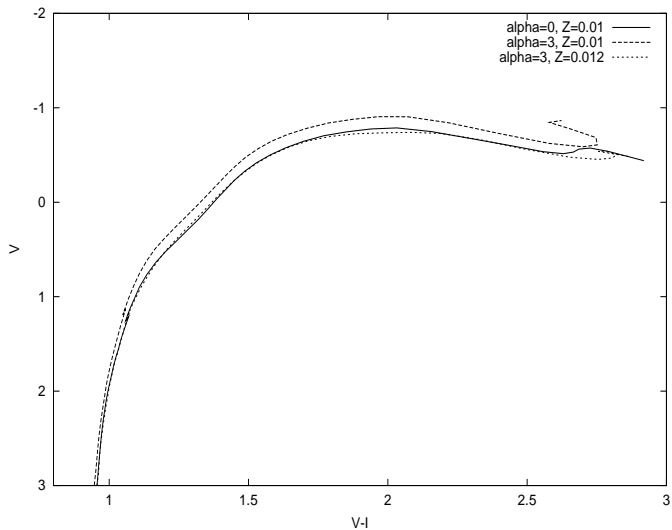}
\caption
{Here we have plotted isochrones for 10 Gyr with metallicity of one half solar
abundance, varying the alpha element ratios to $[\alpha/Fe] = 0.0, 0.3$.
In addition, we have overplotted an isochrone for $[\alpha/Fe] = 0.3$ and
metallicity 20\% higher.  The effects of varying metallicity and
$[\alpha/Fe]$ are very similar.}
\label{alphaz}
\end{figure}

Varying alpha enhancement ratios gives us yet another degeneracy.  In Fig.
\ref{alphaz} it is clear that the variation between alpha element
enhancement of $[\alpha/Fe] = 0.0$ and $0.3$ at an age of 10 Gyr and
metallicity of half-solar, is approximately the same as that caused by a
metallicity increase of 20\%.  This gives us another contribution to the
error on our metallicity measurements.

The systematic effects of incorrectly modelled alpha element abundances can
only have a small differential effect.  For example, in modelling galactic
globular clusters, all clusters may exhibit similar alpha-element ratios,
and thus any error in assuming an incorrect alpha-element ratio for the
isochrones will result in a constant shift for all clusters, retaining
exactly the correct ordering if not the precise metallicity values.

  \subsubsection{Parameter fine-tuning}
  \label{tweak}

With so many parameters, it is usually possible to alter stellar models
in such a way so as to fit an observed population with a reasonable age and
metallicity spread.  Many groups, for example Bergbusch \& VandenBerg (2001)
do exactly this, tuning their isochrones so that they fit a set of template
populations, as well as standard data for the sun.  In this case,
comparisons are made with galactic globular clusters M3 and M92. and the
assumption is that these populations are representative.

In some sense what we are doing here is using the answer as a starting block
for our models, and then we should not be too surprised when our models give
us back the answer upon which we based them.  However, once one moves to
different clusters, and indeed other stellar populations, such as Population
II halo stars, the benefit of this method becomes apparent.  Of course, we
are assuming that there is some underlying law, or set of laws, which unify
all stellar populations given a set of observable parameters.  Observations
show us that this might not necessarily be the case.  It is only an
assumption that models tuned to fit globular clusters should be
preferentially more accurate in analysing stellar haloes.

Even after fitting their models to a wide range of MWG globular clusters,
Bergbusch \& VandenBerg still found that their models failed to predict the
observed luminosities of the red giant bump by $\simeq 0.25 mag$. No matter
how accurately model parameters are optimised, there is always the danger that
we have neglected some fundamental physics which complicates the situation.

 \subsection{Non-linear isochrone interpolation}

When interpolating between isochrones, we are forced to make some
simplifying assumptions.  One such assumption is that a linear interpolation
is sufficiently accurate for our needs.  That is, we assume that e.g.
between two isochrones of metallicity $Z = 0.01$ and $Z = 0.011$ there might
be another 9 isochrones all spaced by 0.0001 in metallicity, $Z$.  However,
the assumption that these interpolated isochrones should lie exactly evenly
spaced in colour-luminosity space is one that might not be valid.

However, when interpolating new isochrones we already have a great deal of
difficulty pairing off the four isochrones spanning our desired
age/metallicity values in both these variables.  To improve on the linear
interpolation assumption would require spline-fitting from previous
isochrones, which would then need to be paired off with the target
isochrones in a similar way.  This method then rapidly becomes unfeasible.

The best guess we can make is that the isochrones do not behave
pathologically and that, in the absence of other information, isochrones
linearly spaced in metallicity should be linearly spaced in the
colour-luminosity variables that we are calculating.

Further interpolation problems lie with the isochrone masses.  Isochrones
are parametrized as a series of points approximating a parametric curve.
However, assuming straight-line sections joining these points can introduce
errors in the assumed masses, and therefore the occupation probabilities for
those points.  Harris \& Zaritsky (2001) investigated this problem, and
discovered two important things;

\begin{itemize}
\item {The error is never more than a few hundredths of a solar mass, and is
therefore fairly insignificant in terms of IMF variability.}
\item {The error is negligible for stars brighter than the MSTO, i.e. all
the stars in which we are interested.}
\end{itemize}

Bergbusch \& VandenBerg (2001) introduce the concept of an Akima spline for
interpolation accurate colours around the MSTO.  Again, these effects are
small, being of the order $\Delta T_{eff} \simeq 0.001 T_{eff}$, and are
confined to low-luminosity stars.

We therefore neglect these effects entirely.

  \subsection{Conclusions}

Isochrone fitting clearly has its limitations, but once these are accepted
and dealt with as effectively as possible, it becomes a powerful method for
analysing stellar populations.  However, one must be aware of the inherent
problems and understand that any results obtained are only useful in a
relative sense; the absolute age and metallicity values are likely to be
highly unreliable in some cases.

Most importantly, the uncertainty in isochrones is most apparent in young,
high mass models where more extreme and poorly understood physics is
operating.  Not surprisingly, models of solar-mass stars are good.  For the
studies which are considered in this work, very few stars are above one
solar mass.  Most populations are of 10 Gyr or older, at which age the
oldest star on the RGB will have an initial mass of only fractionally above
one solar mass for a solar metal abundance (from Girardi et al. isochrones).
The worst of these problems can therefore be safely neglected.

\label{lastpage}

\begin{thebibliography}{}
\bibitem{bell}Bellazzini M., Ferraro F.R., Pancino E., 2001, ApJ, 556, 635
\bibitem{bergbusch}Bergbusch P.A., VandenBerg D.A., 2001, ApJ 556, 322
\bibitem{bertelli}Bertelli G., Betto R., Chiosi C., Bressan A., Nasi E., 1990, A\&AS 85, 845
\bibitem{ba}Bertin E., Arnouts S., 1996, A\&AS, 117, 393
\bibitem{cml}Cioni M.-R.L., van der Marel R.P., Loup C., Habing H.J., 2000, A\&A, 359, 601
\bibitem{dca}Da Costa G.S., Armandroff T.E., 1990, AJ, 100, 162
\bibitem{ferr}Ferrarese L., et al. 1996, ApJ, 464, 568
\bibitem{fg1}Frayn C.M., Gilmore G.F., 2002, MNRAS, In Press
\bibitem{girardi}Girardi L., Bressan A., Bertelli G., Chiosi C., 2000, A\&AS, 141, 371
\bibitem{hz}Harris J., Zaritsky D., 2001, ApJS, 136, 25
\bibitem{harris}Harris W.E., 1996, AJ 112, 1487
\bibitem{kleyna}Kleyna J.T., Geller M.J., Kenyon S.J., Kurtz M.J., Thorstensen J.R., 1998, AJ, 115, 2359
\bibitem{maeder}Maeder A., Meynet G., 1991, A\&ASS, 89, 451
\bibitem{mateo}Mateo M.L., 1998, ARA\&A, 36, 435
\bibitem{mathis}Mathis J.S., 1990, AnnRev., 28, 37
\bibitem{rogers}Rogers F.J., Iglesias C.A., 1996, AAS, 188, 5804
\bibitem{rosenberg}Rosenberg A., Aparicio A., Saviane I., Piotto G., 2000, A\&AS, 144,5
\bibitem{sakai}Sakai S., Madore B.F., Freedman W.L., 1996, ApJ, 461, 713
\bibitem{sc}Salaris M., Cassisi S., 1997, MNRAS, 289, 406
\bibitem{salaris}Salaris M., Chieffi A., Straniero O., 1993, ApJ, 414, 580
\bibitem{scw}Salaris M., Cassisi S., Weiss A., 2002, PASP, 114, 375
\bibitem{yi}Yi S., Demarque P., Kim Y.-C., Lee Y.-W., Ree C.H., Lejeune T.,
Barnes S., 2001, ApJS, 136, 417
\bibitem{yildiz}Yildiz M., Kiziloglu N., 1997, A\&A, 326, 187
\end{thebibliography}
\end{document}